\documentclass[12pt]{article}
\usepackage{amsfonts,amssymb}

\newfont{\bbd}{msbm10 scaled\magstep1}

\def\l#1#2{\raisebox{.2ex}{$\displaystyle
  \mathop{#1}^{{\scriptstyle #2}\widerightarrow}$}}
\def\r#1#2{\raisebox{.2ex}{$\displaystyle
 \mathop{#1}^{\wideleftarrow {\scriptstyle #2}}$}}

\begin{document}
\thispagestyle{empty}

\def\ve#1{\mid #1\rangle}
\def\vc#1{\langle #1\mid}

\newcommand{\p}[1]{(\ref{#1})}
\newcommand{\be}{\begin{equation}}
\newcommand{\ee}{\end{equation}}
\newcommand{\sect}[1]{\setcounter{equation}{0}\section{#1}}


\newcommand{\vs}[1]{\rule[- #1 mm]{0mm}{#1 mm}}
\newcommand{\hs}[1]{\hspace{#1mm}}
\newcommand{\mb}[1]{\hs{5}\mbox{#1}\hs{5}}
\newcommand{\Db}{{\overline D}}
\newcommand{\bea}{\begin{eqnarray}}

\newcommand{\eea}{\end{eqnarray}}
\newcommand{\wt}[1]{\widetilde{#1}}
\newcommand{\und}[1]{\underline{#1}}
\newcommand{\ov}[1]{\overline{#1}}
\newcommand{\sm}[2]{\frac{\mbox{\footnotesize #1}\vs{-2}}
           {\vs{-2}\mbox{\footnotesize #2}}}
\newcommand{\prt}{\partial}
\newcommand{\eps}{\epsilon}

\newcommand{\R}{\mbox{\rule{0.2mm}{2.8mm}\hspace{-1.5mm} R}}
\newcommand{\Z}{Z\hspace{-2mm}Z}

\newcommand{\cd}{{\cal D}}
\newcommand{\cg}{{\cal G}}
\newcommand{\ck}{{\cal K}}
\newcommand{\cw}{{\cal W}}

\newcommand{\vj}{\vec{J}}
\newcommand{\vl}{\vec{\lambda}}
\newcommand{\vz}{\vec{\sigma}}
\newcommand{\vt}{\vec{\tau}}
\newcommand{\vw}{\vec{W}}
\newcommand{\poiss}{\stackrel{\otimes}{,}}

\def\l#1#2{\raisebox{.2ex}{$\displaystyle
  \mathop{#1}^{{\scriptstyle #2}\rightarrow}$}}
\def\r#1#2{\raisebox{.2ex}{$\displaystyle
 \mathop{#1}^{\leftarrow {\scriptstyle #2}}$}}



\renewcommand{\thefootnote}{\fnsymbol{footnote}}
\newpage
\setcounter{page}{0}
\pagestyle{empty}
\begin{flushright}
{ITP-UH-26/06}\\
{JINR-E2-2006-170}
\end{flushright}
\vfill

\begin{center}
{\LARGE {\bf  N=2 supersymmetric unconstrained}}\\[0.3cm]
{\LARGE {\bf matrix GNLS hierarchies are consistent}}\\[1cm]

{}~

{\large F. Delduc$^{a,1}$, O. Lechtenfeld$^{b,2}$, and A.S. Sorin$^{c,3}$}
{}~\\
\quad \\
{\em {~$~^{(a)}$ Laboratoire de Physique$^\dagger$,
Groupe de Physique Th\'eorique,}}\\
{\em ENS Lyon, 46 All\'ee d'Italie, 69364 Lyon, France}\\[10pt]
{\em {~$~^{(b)}$ Institut f\"ur Theoretische Physik, Leibniz Universit\"at
Hannover,}}\\
{\em Appelstra\ss{}e 2, D-30167 Hannover, Germany}\\[10pt]
{\em {~$~^{(c)}$ Bogoliubov Laboratory of Theoretical Physics,}}\\
{\em {Joint Institute for Nuclear Research,}}\\
{\em 141980 Dubna, Moscow Region, Russia}~\quad\\

\end{center}

\vfill

{}~

\centerline{{\bf Abstract}} \noindent
We develop a pseudo--differential approach to the
$N{=}2$ supersymmetric unconstrained matrix $(k|n,m)$--Generalized
Nonlinear Schr\"odinger hierarchies and prove consistency of the corresponding
Lax--pair representation (nlin.SI/0201026). Furthermore, we
establish their equivalence to the integrable hierarchies derived in the super--algebraic
approach of the homogeneously-graded loop superalgebra
$sl(2k+n\vert 2k+m)\otimes C[{\lambda},{\lambda}^{-1}]$ (nlin.SI/0206037).
We introduce an unconventional definition of $N=2$ supersymmetric strictly
pseudo--differential operators so as to close their algebra among themselves.

{}~

{}~



{}~

{}~

\vfill
{\em \noindent
1) E-Mail: francois.delduc@ens-lyon.fr\\
2) E-Mail: lechtenf@itp.uni-hannover.de\\
3) E-Mail: sorin@theor.jinr.ru }\\
$\dagger$) UMR 5672 du CNRS, associ\'ee \`a l'Ecole Normale Sup\'erieure de
Lyon.
\newpage
\pagestyle{plain}
\renewcommand{\thefootnote}{\arabic{footnote}}
\setcounter{footnote}{0}

\section{Introduction and Summary}

The $N{=}2$ supersymmetric unconstrained matrix
$(k|n,m)$--Generalized Nonlinear Schr\"odinger ($(k|n,m)$--GNLS)
hierarchies were proposed in~\cite{ks1} by exhibiting the
corresponding {\it matrix pseudo--differential} Lax--pair
representation
\begin{eqnarray*}
L \ =\ I\partial + F D{\overline D}\partial^{-1} {\overline F}
\end{eqnarray*}
in terms of a $k\times(m{+}n)$ matrix $F$ with
{\it $N{=}2$ unconstrained superfield} entries
for the bosonic isospectral flows.
Their super--algebraic formulation and recursion relations
were proposed in~\cite{Delduc-Sorin} on the basis of the
homogeneously-graded loop superalgebra
$sl(2k+n\vert 2k+m)\otimes C[{\lambda},{\lambda}^{-1}]$.

These hierarchies generalize
and contain as limiting cases many other interesting $N{=}2$
supersymmetric hierarchies discussed in the literature: When
matrix entries are chiral and antichiral $N{=}2$ superfields, these
hierarchies reproduce the $N{=}2$ chiral matrix $(k|n,m)$-GNLS
hierarchies~\cite{bks1,bks2}, and in turn the latter coincide with
the $N{=}2$ GNLS hierarchies of references~\cite{bks,bs} in the
scalar case $k{=}1$. When matrix entries are unconstrained $N{=}2$
superfields and $k{=}1$, these hierarchies are equivalent to the
$N{=}2$ supersymmetric multicomponent hierarchies~\cite{pop}. The
bosonic limit of the $N{=}2$ unconstrained matrix $(k|0,m)$--GNLS
hierarchy reproduces the bosonic matrix NLS equation elaborated
in~\cite{fk} via the $gl(2k+m)/(gl(2k)\times gl(m))$--coset
construction. The $N{=}2$ matrix $(1|1,0)$--GNLS hierarchy is related to
one of three different existing $N{=}2$ supersymmetric KdV
hierarchies -- the $N{=}2$ ${\alpha}{=}1$ KdV hierarchy -- by a
reduction~\cite{pop,ks1,ks2}.

Self-consistency of the Lax--pair representation for the $N{=}2$
supersymmetric unconstrained matrix $(k|n,m)$--GNLS hierarchies
was actually proven in~\cite{ks1} only for the first four flows,
but conjectured for the general case.
The equivalence of their super--algebraic
and pseudo--differential formulations was
established in~\cite{Delduc-Sorin}, but again for the first few flows only.
The present letter completes these proofs.
In Section~2 we develop a pseudo--differential
approach to the $N{=}2$ supersymmetric unconstrained matrix $(k|n,m)$--GNLS
hierarchies in $N{=}2$ superspace and rigorously construct their
Lax--pair representation
\begin{eqnarray*}
{\textstyle{\partial\over\partial t_p}}L\ =\
\bigl[(L^p)_{\oplus}\ ,\ L\bigr]
\end{eqnarray*}
with
\begin{eqnarray*}
{\textstyle{\partial\over\partial t_p}}F\ =\
\Bigl( ({L^{p}})_{\oplus}F\Bigr)
\qquad\textrm{and}\qquad
{\textstyle{\partial\over\partial t_p}}{\overline F}\ =\
-\Bigl({\overline F}  (\r {L^{p}}{})_{\oplus}\Bigr)\ ,
\end{eqnarray*}
where we introduce an unconventional definition of $N=2$ supersymmetric strictly
pseudo--differential operators so as to close their algebra among
themselves\footnote{The precise notation will be explained there.}.
Furthermore, we produce the recursion relations for the corresponding
isospectral flows. As we establish in Section~3, this Lax--pair
representation agrees with the one derived in the super--algebraic approach
of the homogeneously-graded loop superalgebra
$sl(2k+n\vert 2k+m)\otimes C[{\lambda},{\lambda}^{-1}]$.
Thus, we finally prove the conjectured equivalence of the two hierarchies.

Apart from the Lax--pair representation for the isospectral flows
and the recursion relations of these hierarchies,
we presently do not know other characteristic properties
like their (bi)Hamiltonian structures, discrete symmetries etc.,
although part of these are known for some limiting cases.
We hope to address these problems in future.

\section{Pseudo--differential approach}

Our starting point is the Lax operator for the
$N{=}2$ supersymmetric unconstrained matrix $(k|n,m)$--GNLS hierarchies
introduced in~\cite{ks1}\footnote{Note, the
supermatrices $\{F, {\overline F}\}$ are re-scaled by $\sqrt{2}$
comparing to \cite{ks1}.}
\begin{eqnarray}
L = I\partial + F D{\overline D}\partial^{-1} {\overline F}.
\label{suplax-def}
\end{eqnarray}
Here, $F\equiv
F_{Aa}(Z)$ and ${\overline F}\equiv {\overline F}_{aA}(Z)$
($A,B=1,\ldots, k$; $a,b=1,\ldots , n+m$) are rectangular matrices
which entries are unconstrained $N=2$ superfields, $I$ is the
unity matrix, $I_{AB}\equiv {\delta}_{AB}$, and the matrix product
is implied, for example $(F\overline F)_{AB} \equiv \sum_{a=1}^{n+m}
F_{Aa}\overline F_{aB}$ and $(F\overline F)_{ab} \equiv \sum_{A=1}^{k}
F_{aA}\overline F_{Ab}$. The matrix entries are Grassmann even
superfields for $a=1,\ldots ,n$ and Grassmann odd superfields for
$a=n+1,\ldots , n+m$. Thus, fields do not commute, but rather
satisfy $F_{Aa}{\overline F}_{bB}=(-1)^{d_{a}{\overline d}_{b}}
{\overline F}_{bB}F_{Aa}$ where $d_{a}$ and ${\overline d}_{b}$
are the Grassmann parities of the matrix elements $F_{Aa}$ and
${\overline F}_{bB}$, respectively, $d_{a}=1$ $(d_{a}=0)$ for odd
(even) entries. Superfields depend on the coordinates
$Z=(z,\theta,\overline\theta)$ of $N=2$ superspace, and
the $N=2$ supersymmetric fermionic covariant derivatives
$D,{\overline D}$ are
\begin{eqnarray}
D=\frac{\partial}{\partial\theta}
 -\frac{1}{2}\overline\theta\frac{\partial}{\partial z}, \quad
{\overline D}=\frac{\partial}{\partial\overline\theta}
 -\frac{1}{2}\theta\frac{\partial}{\partial z}, \quad
D^{2}={\overline D}^{2}=0, \quad
\left\{ D,{\overline D} \right\}= -\frac{\partial}{\partial z}
\equiv -{\partial}.
\label{DD}
\end{eqnarray}

The set of operators
\begin{equation}
\{{\hat o}_i, i\in \hbox{\bbd Z}\}:=
\{D^{n}{\overline D}^{\overline n}{\partial}^m,~
n, \overline n = 0,1, ~ m\in \hbox{\bbd Z}\}\label{basisel}
\end{equation}
 forms a basis in the associative
algebra of the supermatrix valued pseudo--differential operators on $N=2$ superspace
\begin{eqnarray}
{\hbox{\bbd O}}=\sum^{\infty}_{i=-\infty}f_{i} {\hat o}_i:=
\sum^{N_{max}}_{m=-\infty}\sum^{1}_{n,\overline n=0}f_{n,\overline n, m}
D^{n}{\overline D}^{\overline n}{\partial}^m, \quad
d_{{\hbox{\bbd O}}}=d_{\hat o_i}+d_{f_i}
\label{pseudo-dif-canon}
\end{eqnarray}
where $f_i$ is a supermatrix valued $N=2$ superfield and ${\hat o}_i$ is a
basis operator with the Grassmann parities $d_{f_i}$ and $d_{{\hat o}_i}$,
respectively, and we understand that the operator ${\hbox{\bbd O}}$
possesses a definite Grassmann parity $d_{{\hbox{\bbd O}}}$.
We shall say that {\bbd O} above is a differential operator if the sum over $m$ is restricted
to positive or zero values only, and that {\bbd O} is strictly pseudo--differential
if the sum over $m$ is restricted to negative values of $m$. The set of the differential
operators and the set of the strictly pseudo--differential operators both form a subalgebra
of the whole space of pseudo--differential operators. Any pseudo--differential operator {\bbd O}
is the sum of a differential operator $\hbox{\bbd O}_{\oplus}$ and a strictly pseudo--differential
operator $\hbox{\bbd O}_{\ominus}$. Hereafter,  the notation $(L^p)_{\hat o_i}$ denotes the
supermatrix coefficient of the basis element ${\hat o_i}$, i.e. $(L^p)_{\hat o_i}{\hat o_i}$
belongs to the expansion (\ref{pseudo-dif-canon}) of $L^p$;
$(\hbox{\bbd O}f)$ has the meaning of a supermatrix valued pseudo--differential operator
$\hbox{\bbd O}$ acting only on a supermatrix valued function $f$ inside
the brackets\footnote{In order to avoid misunderstanding,
let us remark the difference in the notations $(\hbox{\bbd O}f)$ and $\{(\hbox{\bbd O}f)_{\oplus},
~(\hbox{\bbd O}f)_{\ominus}\}$:
the former represents a supermatrix valued superfield, while the latter corresponds
to a differential and strictly pseudo--differential parts of the operator $\hbox{\bbd O}f$,
respectively.}.

{\underline {\bf Remark.}}
The definition which is used in this paper of a strictly pseudo--differential operator slightly differs
from the one which was used in the articles \cite{ks1,Delduc-Sorin}. There, instead of the basis elements
$D\overline D\partial^n$, one was rather using as basis elements the operators $[D,\overline D]\partial^n$.
Differential and strictly pseudo--differential operators were defined with respect to this basis.
Any pseudo--differential operator {\bbd O} may be separated into a differential operator
${\hbox{\bbd O}}_{+}$ and a strictly pseudo--differential operator ${\hbox{\bbd O}}_{-}$
according to this basis. This other definition has the drawback that strictly pseudo--differential
operators do not form a closed algebra, because of the relation $([D,\overline D]\partial^{-1})^2=1$.
The relation between the two definitions is easily obtained by using the relation
$D\overline D\partial^{-1}=-\frac{1}{2}+\frac{1}{2}[D,\overline D]\partial^{-1}$.
Defining the residue of the operator {\bbd O} as the coefficient of  $D\overline D\partial^{-1}$
in the new basis (this differs by a factor of $2$ from the definition in \cite{ks1,Delduc-Sorin},
where it was defined as the coefficient of $[D,\overline D]\partial^{-1}$), one finds the following relation
\begin{equation}{\hbox{\bbd O}}_{\oplus}={\hbox{\bbd O}}_{+}+\frac{1}{2}res{\hbox{\,\bbd O}},
\end{equation}
which allows one to relate calculations in previous articles and in the present article.

{\underline {\bf Definition 1.}}
We define the involutive automorphism $^*$ of the second order of the supersymmetry algebra
\begin{eqnarray}
\{\partial,~ D, ~\overline D\}^*=-\{\partial,~ D, ~\overline D\}, \quad
(D \overline D)^{*}=-\overline D D, \quad (\overline D D)^{*}=-D \overline D.
\label{involution}
\end{eqnarray}
It can be extended to all basis elements in (\ref{basisel}) using the rule
$({\hat o}_i {\hat o}_j)^{*}=(-1)^{d_{{\hat o}_i}d_{{\hat o}_j}}{\hat o}_j^*{\hat o}_i^*$.
When applied to a supermatrix $f\Rightarrow f^*$ simply amounts to a change
in the sign of its Grassmann--odd entries. Its k-fold action on the
supermatrix $f$ will be denoted $f^{*(k)}$ ($k= 0,1~ mod ~2$).

{\underline {\bf Remark.}}
Relations \p{involution} reproduce the conventional operator-conjugation rules for
the fermionic and bosonic covariant derivatives, although the star--operation $^*$,
being applied to a supermatrix $f^*$, differs comparing to the conventional operation
of the super--transposition of a supermatrix $f^T$, i.e. $f^*\neq f^T$. We also remark
that $(F\overline F)^*\equiv F^*{\overline F}^* =F\overline F$, $F{\overline F}^*=F^* \overline F$,
while $(\overline F F)^*\equiv {\overline F}^* F^* \neq \overline F F$.

{\underline {\bf Definition 2.}}
We introduce the adjoint operator $\r {\hbox{\bbd O}}{}$ by defining  its
action on the supermatrix valued superfield $f$ with the Grassmann parity $d_f$
\begin{eqnarray}
f ~\r {\hbox{\bbd O}}{}:=\sum^{\infty}_{i=-\infty}
{\hat o}^{*}_i (f f_i)^{*(d_{{\hat o}_i})}.
\label{def2}
\end{eqnarray}

{\underline {\bf Remark.}}
This definition generalizes the definition of the adjoint operator to the non-abelian,
noncommutative case. For the abelian, commutative case, i.e. when $f$ and $f_i$
are not (super)matrices, but commutative functions, it reproduces the conventional
definition of the adjoint operator. Due to this reason we call the operator
$\r {\hbox{\bbd O}}{}$ noncommutatively--adjoint operator.

Equation \p{def2} defines a product of a noncommutatively--adjoint operator
and a supermatrix--valued superfield. In order to consistently define a product
of different noncommutatively--adjoint operators with themselves, we firstly need
to prove:

{\underline {\bf Proposition 1.}}
\begin{eqnarray}
&&(f~\overleftarrow{{\hbox{\bbd O}_{1}}...{\hbox{\bbd O}_{k}}})
=(f~\r {\hbox{\bbd O}}{}_{1}...\r {\hbox{\bbd O}}{}_{k})
\equiv(((f~\r {\hbox{\bbd O}}{}_{1})...)\r {\hbox{\bbd O}}{}_{k}).
\label{prop-1}
\end{eqnarray}

{\underline {\bf Proof.}} By induction, it is sufficient to check (\ref{prop-1}) for two operators
${\hbox{\bbd O}}{}_{1}$ and ${\hbox{\bbd O}}{}_{2}$, which has been done with the help of the rules
(\ref{involution},\ref{def2}).
$~~~~~\blacksquare$

{\underline {\bf Remark.}} Proposition 1 gives the result of the action of a product
of noncommutatively-adjoint operators on a supermatrix--valued superfield. Using
the latter we define a product $\prod_{i=1}^{k} \r {\hbox{\bbd O}}{}_i$ of
noncommutatively--adjoint operators which generalizes the definition of the
product of the conjugated operators in the commutative case to the noncommutative
case.

{\underline {\bf Definition 3.}}
\begin{eqnarray}
f~\r {\hbox{\bbd O}}{}_{1}...\r {\hbox{\bbd O}}{}_{k}
:=f~\overleftarrow{{\hbox{\bbd O}_{1}}...{\hbox{\bbd O}_{k}}}.
\label{prop-def}
\end{eqnarray}

{\underline {\bf Remark.}}
It is obvious that the r.h.s. of eq. \p{prop-def} can be calculated using
eq. \p{def2}, if one takes into account that due to the associativity of the
algebra of pseudo--differential operators the product
${\hbox{\bbd O}}{}_{1}...{\hbox{\bbd O}}_{k}={\hbox{\bbd O}}$,
where ${\hbox{\bbd O}}$ is a pseudo--differential operator in the canonical form
\p{pseudo-dif-canon}, therefore one can use \p{def2}. The consistency of eq. \p{prop-def} with
eq. \p{def2} is provided by eq. \p{prop-1}.

{\underline {\bf Lemma 1.}}
\begin{eqnarray}&&
D\overline D\partial^{-1}fD\overline D\partial^{-1}=(D\overline D\partial^{-1}f)D\overline D\partial^{-1}
+D\overline D\partial^{-1}(f\overleftarrow{D\overline D\partial^{-1}}),\label{fi}\\&&
({\hbox{\bbd O}}_{\oplus}fD\overline D\partial^{-1})_{\ominus}=({\hbox{\bbd O}}_{\oplus}f)D\overline D\partial^{-1},
\label{se}\\&&
(D\overline D\partial^{-1}f{\hbox{\bbd O}}_{\oplus})_{\ominus}
=D\overline D\partial^{-1}(f\overleftarrow{{\hbox{\bbd O}}_{\oplus}}).\label{th}
\end{eqnarray}

{\underline {\bf Proof.}} Equality (\ref{fi}) results from the following simple relation :
\begin{equation}
\label{prooflemma1}
\partial^{-1}f\partial^{-1}=(\partial^{-1}f)\partial^{-1}+
\partial^{-1}(f\overleftarrow{\partial^{-1}})
\end{equation}
One acts with $D\overline D$ on both sides of (\ref{prooflemma1}), then tries to push $D$ and $\overline D$
to the right in the first term, and to the left in the second term.

Equality \p{se} is obvious if one takes into account that the pseudo--differential operator
$D{\overline D}\partial^{-1}$ being multiply either by $D$ or ${\overline D}$ from the right
or left becomes a differential operator. Equality \p{th} is an operator-adjoint counterpart of
equality \p{se}. $~~~~~\blacksquare$

It should be noted that, although there is an arbitrariness in the definition of the action
of $\partial^{-1}$ on a function $f$, this arbitrariness does not show up in (\ref{fi}) because
it compensates between both terms on the right-hand side.

{\underline {\bf Proposition 2.}}
\begin{eqnarray}
(L^p)_{\ominus} = \sum^{p-1}_{k=0}
(L^{p-k-1}F) D{\overline D}\partial^{-1}
({\overline F}~ \overleftarrow{L^{k}}{}), \quad p \in \hbox{\bbd N}.
\label{laxmp}
\end{eqnarray}

{\underline {\bf Proof.}} The proof is by induction with the use of relations (\ref{fi}--\ref{th}).
Equation (\ref{laxmp}) is obviously correct at $p=1$ (compare with eq. (\ref{suplax-def})).
If it is correct for the $p=n$ case, then we have for the $p=n+1$ case
\begin{eqnarray}
(L^{n+1})_{\ominus}&\equiv& (LL^n)_{\ominus}\cr
&=&\left(\Big(I\partial+FD\overline D\partial^{-1}\overline F\Big)\Big((L^n)_{\oplus}+\sum^{n-1}_{k=0}
(L^{p-k-1}F) D{\overline D}\partial^{-1}
({\overline F} ~\overleftarrow{L^{k}}{})\Big)\right)_{\ominus}\cr
&=&\Big(FD\overline D\partial^{-1}\overline F(L^{n})_{\oplus}
+\sum^{n-1}_{k=0}(\partial L^{n-k-1}F)D\overline D\partial^{-1}
(\overline F~\overleftarrow{L^k})\cr&+&
FD\overline D\partial^{-1}\overline F\sum^{n-1}_{k=0}(L^{n-k-1}F)D\overline D\partial^{-1}
(\overline F~\overleftarrow{L^k})\Big)_{\ominus}\cr&=&
FD\overline D\partial^{-1}\Big(\overline F~\overleftarrow{(L^{n})_{\oplus}}
+\sum^{n-1}_{k=0}\overline F(L^{n-k-1}F)
~\overleftarrow{D\overline D\partial^{-1}}(\overline F~\overleftarrow{L^k})\Big)\cr&+&
\sum^{n-1}_{k=0}\Big((\partial+FD\overline D\partial^{-1}\overline F)L^{n-k-1}F\Big)D\overline D\partial^{-1}
(\overline F~\overleftarrow{L^k})\cr&=&
FD\overline D\partial^{-1}(\overline F~\overleftarrow{L^{n}})+\sum^{n-1}_{k=0}(L^{n-k}F)
D\overline D\partial^{-1}(\overline F~\overleftarrow{L^k})\cr&=&
\sum^{n}_{k=0}(L^{n-k}F)D\overline D\partial^{-1}(\overline F\overleftarrow{L^k}).
\end{eqnarray}
$~~~~~\blacksquare$

{\underline {\bf Proposition 3.}}
\begin{eqnarray}
\label{lax^n-iden1}
(L^{p+1})_{\oplus}&=&
(L^p)_{\oplus}L-((L^p)_{\oplus}F)D\overline D\partial^{-1}\overline F+\sum^{p-1}_{k=0}
(L^{p-k-1}F) D{\overline D}(\overline F~\overleftarrow{L^k}{})\label{p1}\\ &=&
L(L^p)_{\oplus}-FD\overline D\partial^{-1}(\overline F~\overleftarrow{(L^p)_{\oplus}})
+\sum^{p-1}_{k=0}(L^{p-k-1}F) D{\overline D}
({\overline F}~\overleftarrow{L^{k}}{}).
\label{1p}
\end{eqnarray}

{\underline {\bf Proof.}
This is an easy calculation using eqs. (\ref{se},\ref{th},\ref{laxmp}) and obvious identities
\begin{eqnarray}
\label{identities-new}
(L^{p+1})_{\oplus}&=&
(L^p)_{\oplus}L-((L^p)_{\oplus}L_{\ominus})_{\ominus}+((L^p)_{\ominus}L_{\oplus})_{\oplus}
\cr &=&
L(L^p)_{\oplus}-(L_{\ominus}(L^p)_{\oplus})_{\ominus}+(L_{\oplus}(L^p)_{\ominus})_{\oplus}.
\end{eqnarray}
$~~~~~\blacksquare$

{\underline {\bf Corollary.}}
Subtracting eq. \p{1p} from eq. \p{p1} we obtain
\begin{eqnarray}
[(L^p)_{\oplus},L]=((L^p)_{\oplus}F)
D{\overline D}\partial^{-1}{\overline F}
-FD{\overline D}\partial^{-1}
({\overline F} (\r {L^{p}}{})_{\oplus}).
\label{ident1}
\end{eqnarray}

If one introduces evolution derivatives (flows)
${\textstyle{\partial\over\partial t_p}}$ according to the formula
\begin{eqnarray}
{\textstyle{\partial\over\partial t_p}}F= (({L^{p}})_{\oplus}F), \quad
{\textstyle{\partial\over\partial t_p}}{\overline F}=
-({\overline F} ((\r {L^{p}}{})_{\oplus} ),
\label{evol-eq}
\end{eqnarray}
then eq. \p{ident1} takes the form of the Lax pair representation
\begin{eqnarray}
{\textstyle{\partial\over\partial t_p}}L =[(L^p)_{\oplus},L]
\label{suplax}
\end{eqnarray}
which was proposed in \cite{ks1}. Actually, its self-consistency was proven
in \cite{ks1} only for the first few flows p=0,1,2
and 3, then conjectured for the general case there, and the corresponding
integrable hierarchies were called the $N=2$ supersymmetric unconstrained matrix
$(k|n,m)$--Generalized Nonlinear Sch\"odinger hierarchies.
The algebra of the flows in \p{suplax} can easily be calculated
\begin{eqnarray}
[{\textstyle{\partial\over\partial t_m}},
{\textstyle{\partial\over\partial t_n}}]=0,
\label{alg}
\end{eqnarray}
it is abelian algebra of the isospectral flows.
The Lax--pair representation \p{suplax} may be seen as the
integrability condition for the following linear system:
\begin{eqnarray}
L\psi_1  &=& {\lambda} \psi_1, \label{spectrstart} \\
{\textstyle{\partial\over\partial t_p}}\psi_1 &=& ((L^{p})_{\oplus} \psi_1)
\label{spectr}
\end{eqnarray}
where $\lambda$ is the spectral parameter and the eigenfunction $\psi_1$
is the Baker-Akhiezer function of the hierarchy.

Projecting the Lax--pair representation \p{suplax} on $D{\overline D}\partial^{-1}$,
$D\partial^{-1}$, ${\overline D}\partial^{-1}$ and $\partial^{-1}$ parts, one can
straightforwardly extract
the following evolution equations
\begin{eqnarray}
\label{fbarf}
({\textstyle{\partial\over\partial t_p}}F{\overline F}) &=&
(res(L^p))~' +[res(L^p), F{\overline F}],\\
\label{fbardbarf}
({\textstyle{\partial\over\partial t_p}}F^*{\overline D}~{\overline F})
&=& -(L^p)^{~'}_{D\partial^{-1}} + F{\overline F}~(L^p)_{D\partial^{-1}}
+res(L^p)~(F^*{\overline D}~{\overline F}),\\
\label{fdbarf}
({\textstyle{\partial\over\partial t_p}}F^*D{\overline F})
&=&(L^p)^{~'}_{{\overline D}\partial^{-1}}
+(L^p)_{{\overline D}\partial^{-1}}F{\overline F}
-(F^*D{\overline F})~res(L^p)\nonumber\\
&-&F{\overline F}(D~res(L^p))+res(L^p)(D F{\overline F}),\\
({\textstyle{\partial\over\partial t_p}}FD{\overline D}~{\overline F})
&=&(L^p)^{~'}_{\partial^{-1}}
+(F^* D {\overline F}(L^p)_{D\partial^{-1}})\nonumber\\
&+&(L^p)_{{\overline D}\partial^{-1}}(F^*{\overline D}~{\overline F})
+ res(L^p)(D F^*{\overline D}~{\overline F})
\label{fdbardbarf}
\end{eqnarray}
which can be used to express $res(L^p)$,
$(L^p)_{D\partial^{-1}}$, $(L^p)_{{\overline D}\partial^{-1}}$
and $(L^p)_{\partial^{-1}}$,
entering these equations, in terms of the time derivative
${\textstyle{\partial\over\partial t_p}}$ of different functionals of
$F$ and ${\overline F}$. With this aim we need to introduce a $k\times k$
matrix $g$ by the consistent set of equations.

{\underline {\bf Definition 4.}}
\begin{eqnarray}
g~'=-gF{\overline F}, \quad
(Dg)=-\Bigl({\partial}^{-1}g(D F {\overline F})g^{-1}\Bigr)g, \quad
({\overline D}g)=-\Bigl({\partial}^{-1}g({\overline D}F
{\overline F})g^{-1}\Bigr)g.
\label{def}
\end{eqnarray}

With the help of $g$ the resolution of eqs. (\ref{fbarf}--\ref{fdbardbarf}) with
respect to $res(L^p)$, $(L^p)_{D\partial^{-1}}$,
$(L^p)_{{\overline D}\partial^{-1}}$ and
$(L^p)_{\partial^{-1}}$
is rather simple
\begin{eqnarray}
\label{d-dbar}
res(L^p)&=&-(g^{-1}{\textstyle{\partial\over\partial t_p}}g)
\equiv (\partial^{-1}{\textstyle{\partial\over\partial t_p}}
F {\overline F}g^{-1})g,\\
\label{d}
(L^p)_{D\partial^{-1}}&=& -g^{-1}(\partial^{-1}{\textstyle{\partial\over\partial t_p}}
g F^* {\overline D}~{\overline F}) \equiv
-(\partial^{-1}{\textstyle{\partial\over\partial t_p}}F^*
{\overline D}~{\overline F})\nonumber\\
&+&(\partial^{-1}{\textstyle{\partial\over\partial t_p}}F{\overline F}
g^{-1})(\partial^{-1}g F^* {\overline D}~{\overline F})-
(\partial^{-1}{\textstyle{\partial\over\partial t_p}}F{\overline F}
g^{-1}\partial^{-1}g F^* {\overline D}~{\overline F}),~~~~\\
\label{d-bar}
(L^p)_{{\overline D}\partial^{-1}}&=&
[(\partial^{-1}{\textstyle{\partial\over\partial t_p}}F^* D{\overline F}
g^{-1}) g],\\
(L^p)_{\partial^{-1}}&=&
(\partial^{-1}{\textstyle{\partial\over\partial t_p}}F D{\overline D}
~{\overline F}) -[(\partial^{-1}{\textstyle{\partial\over\partial t_p}}F^* D{\overline F}
g^{-1})(\partial^{-1}g F^* {\overline D}~{\overline F})]\nonumber\\
&+& (\partial^{-1}{\textstyle{\partial\over\partial t_p}}F^* D{\overline F}
g^{-1}\partial^{-1}g F^* {\overline D}~{\overline F})
\label{const-part}
\end{eqnarray}
where in eqs. (\ref{d-bar}--\ref{const-part}) the fermionic derivative $D$ entering
the square brackets acts on the right inside these brackets. This can easily
be verified by directly substituting these expressions into the original
equations (\ref{fbarf}--\ref{fdbardbarf}) and using eqs. \p{def}.

{\underline {\bf Proposition 4.}}
\begin{eqnarray}
\label{lax^n-iden5}
&&(L^{p+1})_{\oplus}\nonumber\\
&&= (L^p)_{\oplus} L-
({\textstyle{\partial\over\partial t_p}}F)D{\overline D}\partial^{-1}
{\overline F}
+(\partial^{-1}{\textstyle{\partial\over\partial t_p}}F D{\overline D}
~{\overline F}) \nonumber\\
&&-(\partial^{-1}{\textstyle{\partial\over\partial t_p}}F^* D{\overline F}
g^{-1})(\partial^{-1}g F^* {\overline D}~{\overline F})+
(\partial^{-1}{\textstyle{\partial\over\partial t_p}}F^* D{\overline F}
g^{-1}\partial^{-1}g F^* {\overline D}~{\overline F})\\
&&=L (L^p)_{\oplus}+
FD{\overline D}\partial^{-1}
({\textstyle{\partial\over\partial t_p}}{\overline F})
+(\partial^{-1}{\textstyle{\partial\over\partial t_p}}F D{\overline D}
~{\overline F}) \nonumber\\
&&-(\partial^{-1}{\textstyle{\partial\over\partial t_p}}F^* D{\overline F}
g^{-1})(\partial^{-1}g F^* {\overline D}~{\overline F})+
(\partial^{-1}{\textstyle{\partial\over\partial t_p}}F^* D{\overline F}
g^{-1}\partial^{-1}g F^* {\overline D}~{\overline F})
\label{lax^n-iden6}
\end{eqnarray}
where in these equations the fermionic derivatives $D$ and ${\overline D}$
entering the brackets act as operators on the right both inside and outside
the brackets.

{\underline {\bf Proof.}}
Taking $(L^{p+1})_{\oplus}$ in  \p{lax^n-iden1} and using eqs. \p{evol-eq}
as well as the identity
\begin{equation}
\sum^{p-1}_{k=0}
(L^{p-k-1}F) D{\overline D}(\overline F\r{L^k}{})
=res(L^p)D{\overline D}+(L^p)_{D \partial^{-1}}D
+(L^p)_{{\overline D}\partial^{-1}}{\overline D}
+(L^p)_{\partial^{-1}}
\end{equation}
which follows from eq. \p{laxmp},
one can easily obtain the following expression:
\begin{eqnarray}
(L^{p+1})_{\oplus}&=& ((L^{p})_{\oplus})L
-({\textstyle{\partial\over\partial t_p}}F)
D{\overline D}\partial^{-1}{\overline F}
+res(L^p)~D{\overline D}
\nonumber\\
&+&(L^p)_{D \partial^{-1}}D
+(L^p)_{{\overline D}\partial^{-1}}{\overline D}
+(L^p)_{\partial^{-1}}.
\label{proof4-1}
\end{eqnarray}
Substituting $res(L^p)$ \p{d-dbar}, $(L^p)_{D\partial^{-1}}$ \p{d},
$(L^p)_{{\overline D}\partial^{-1}}$ \p{d-bar}
and $(L^p)_{\partial^{-1}}$
\p{const-part} into eq. \p{proof4-1},
we arrive at the first equality \p{lax^n-iden5}.
The second equality \p{lax^n-iden6} can obviously be derived from
the first one if one substitutes $(L^p)_{\oplus}L$ by
$L (L^p)_{\oplus}+ {\textstyle{\partial\over\partial t_p}} L$
there, according to eq. \p{suplax}.
$~~~~~\blacksquare$

{\underline {\bf Corollary: recursion relations.}}
Applying the noncommutatively--adjoint of operator relation \p{lax^n-iden5} to the
supermatrix valued superfield ${\overline F}$ from the right and similarly
applying \p{lax^n-iden6} to $F$ from the left as well as using eqs. \p{evol-eq}
it is not complicated to obtain recurrence relations relating flows with the evolution
derivatives ${\textstyle{\partial\over\partial t_{p+1}}}$ and
${\textstyle{\partial\over\partial t_{p}}}$
\begin{eqnarray}
\label{rec-rel1}
&&({\textstyle{\partial\over\partial t_{p+1}}}{\overline F})
= -({\textstyle{\partial\over\partial t_{p}}}{\overline F})~'
+({\overline D}D\partial^{-1}{\textstyle{\partial\over\partial t_p}}{\overline F}F)
{\overline F} -[{\overline F}(\partial^{-1}{\textstyle{\partial\over\partial t_p}}
F \r D{}\r {\overline D}{}~{\overline F}) \nonumber\\
&&-{\overline F}(\partial^{-1}{\textstyle{\partial\over\partial t_p}}F^*\r D{}{\overline F}
g^{-1})(\partial^{-1}g F^*\r {\overline D}{}{\overline F})+
{\overline F}(\partial^{-1}{\textstyle{\partial\over\partial t_p}}F^*\r D{}{\overline F}
g^{-1}\partial^{-1}g F^*\r {\overline D}{}{\overline F})],\\
&&({\textstyle{\partial\over\partial t_{p+1}}}F)=
({\textstyle{\partial\over\partial t_{p}}}F)~'+
(FD{\overline D}\partial^{-1}{\textstyle{\partial\over\partial t_p}}{\overline F}F)
+[(\partial^{-1}{\textstyle{\partial\over\partial t_p}}F D{\overline D}
~{\overline F})F \nonumber\\
&&-(\partial^{-1}{\textstyle{\partial\over\partial t_p}}F^* D{\overline F}
g^{-1})(\partial^{-1}g F^* {\overline D}~{\overline F})F+
(\partial^{-1}{\textstyle{\partial\over\partial t_p}}F^* D{\overline F}
g^{-1}\partial^{-1}g F^* {\overline D}~{\overline F}) F]
\label{rec-rel2}
\end{eqnarray}
where the fermionic derivatives $D$ and ${\overline D}$, entering
the square brackets in eqs. \p{rec-rel1} and \p{rec-rel2},
act inside these brackets on the left and right, respectively.

\section{Super--algebraic approach}
Following the super--algebraic
approach{\footnote {For more recent development of the super--algebraic approach,
see ref. \cite{agz-recent} and references therein.}} of ref. \cite{agnpz},
in \cite{Delduc-Sorin} a wide class of integrable hierarchies was constructed
which corresponds to the homogeneous gradation of the loop superalgebra
$sl(2k+n\vert 2k+m)\otimes C[{\lambda},{\lambda}^{-1}]$
with the splitting matrix $E$ and the grading operator $d$,
\begin{eqnarray}
E=\left(\begin{array}{ccccc}
1 &   0 &  0 &  0 &  0 \\
0 &   1 &  0 &  0 &  0 \\
0 &   0 &  0 &  0 &  0 \\
0 &   0 &  0 &  1 &  0 \\
0 &   0 &  0 &  0 &  1
\end{array}\right), \quad d={\lambda}{\textstyle{\partial\over\partial {\lambda}}}.
\label{grad}
\end{eqnarray}
The corresponding isospectral flows are \cite{Delduc-Sorin}
\begin{eqnarray}
\label{locallaxrepr}
({\textstyle{\partial\over\partial t_{p}}}{{\cal L}}_z) =
[ ({\Theta}{\lambda}^p E {\Theta}^{-1})_{+}- (G^{-1}{\textstyle{\partial\over\partial t_{p}}}G) ,
{\cal L}_z ]
\end{eqnarray}
where the dressing matrix $\Theta$ is obtained from dressing the Lax operator ${\cal L}_z$
\begin{eqnarray}
{\cal L}_z
:={\partial}-{\lambda}E + {\cal A}= {{\Theta}}^{-1}\Bigl({\partial}-{\lambda}E\Bigr){\Theta},
\quad {\Theta}= 1 + \sum_{k=1}^{\infty}{\lambda}^{-k} \theta^{(-k)}.
\label{locallaxrepr1}
\end{eqnarray}
Hereafter, the subscript $+$ denotes the projection on the positive homogeneous grading
\p{grad},
\begin{eqnarray}
{\cal A} = \left(\begin{array}{ccccc}
0 &   0 &  F &  0 &  0 \\
0 &   0 &  (DF) &  0 &  0 \\
-(D{\overline D}~{\overline F}) &
({\overline D}~{\overline F}^*) &
0 &  -(D{\overline F}^*) &  -{\overline F} \\
-(F^*{\overline D}~{\overline F}) &
0 &  ({\overline D}F) &
-F{\overline F} & 0 \\
-(DF^*{\overline D}~{\overline F}) &
(F^*{\overline D}~{\overline F}) &
(D{\overline D}F) & -(DF{\overline F}) &
-F{\overline F}
\end{array}\right)
\label{laxmatrixz}
\end{eqnarray}
and
\begin{eqnarray}
G=\left(\begin{array}{ccccc}
1 &   0 &  0 &  0 &  0 \\
0 &   1 &  0 &  0 &  0 \\
0 &   0 &  1 &  0 &  0 \\
-({\partial}^{-1}gF^*{\overline D}~{\overline F}) &   0 &  0 &  g &  0 \\
-(D{\partial}^{-1}gF^*{\overline D}~{\overline F}) &
({\partial}^{-1}gF^*{\overline D}~{\overline F}) &  0 &  (Dg) &  g
\end{array}\right).
\label{transf}
\end{eqnarray}

It is easily seen that the matrix $G$ \p{transf} entering into the Lax--pair representation
\p{locallaxrepr} is nonlocal. Moreover, the $N=2$ superfield entries of the
connection $\cal A$ \p{laxmatrixz} are not independent quantities,
i.e. they are subjected to constraints. Why in this case do
isospectral matrix flows \p{locallaxrepr} be local, as it is
obviously the case for the flows \p{suplax}? Why are they supersymmetric,
or in other words, why do these flows preserve the above--mentioned
constraints? Finally, how can one see in general that these flows
coincide with the isospectral flows \p{suplax}. These
questions were raised in \cite{Delduc-Sorin}, but clarified only partly there.
Based on the pseudo--differential approach developed in the previous Section
we are able to prove here that the super--algebraic isospectral matrix flows \p{locallaxrepr}
are equivalent to the pseudo--differential isospectral flows \p{suplax}, therefore
the former are local and supersymmetric as well, because it is the case for the latter.

The Lax--pair representation \p{locallaxrepr} may be seen as the
integrability condition for the following linear system:
\begin{eqnarray}
\label{psi-eq-1a}
&&{\cal L}_z \Psi = 0, \\
&&({\textstyle{\partial\over\partial t_{p}}}G{\Psi}) =
(G{\Theta}{\lambda}^p E (G{\Theta})^{-1})_{+}G\Psi
\label{psi-eq-2b}
\end{eqnarray}
where $\Psi^T =\left(\psi_1, \psi_2, \psi_3, \psi_4, \psi_5\right)$.
In order to prove equivalence of the Lax--pair representations \p{locallaxrepr}
and \p{suplax} it is enough to prove equivalence of the corresponding linear systems
(\ref{psi-eq-1a}--\ref{psi-eq-2b}) and (\ref{spectrstart}--\ref{spectr}).

{\underline {\bf Proposition 5.}}
The linear systems (\ref{psi-eq-1a}--\ref{psi-eq-2b}) and (\ref{spectrstart}--\ref{spectr})
are equivalent.

{\underline {\bf Proof.}}
The first equation \p{psi-eq-1a} of the linear system (\ref{psi-eq-1a}--\ref{psi-eq-2b})
is equivalent to the first equation \p{spectrstart}
of the linear system (\ref{spectrstart}--\ref{spectr}) and possesses the solution
\begin{eqnarray}
\Psi =
\left(\begin{array}{c}
\psi_1 \\
(D \psi_1) \\
(D\overline D {\partial}^{-1}{\overline F}\psi_1) \\
(\overline D \psi_1) \\
(D\overline D \psi_1)
\end{array}\right)
\label{spectrferm}
\end{eqnarray}
which was actually observed in \cite{Delduc-Sorin}, and it was the starting point for
the super--algebraic construction developed there.

In order to demonstrate that the second equation \p{psi-eq-2b} of the linear system
(\ref{psi-eq-1a}--\ref{psi-eq-2b}) is equivalent to
the second equation \p{spectr} of the linear system (\ref{spectrstart}--\ref{spectr}) as well,
we use the equality \cite{agnpz}
\begin{eqnarray}
\label{rec-id-1}
(G{\Theta}{\lambda}^{p+1} E (G{\Theta})^{-1})_{+}
={\lambda}(G{\Theta}{\lambda}^{p} E (G{\Theta})^{-1})_{+}
+({\textstyle{\partial\over\partial t_{p}}}G{\theta}^{(-1)})
\end{eqnarray}
and rewrite eq. \p{psi-eq-2b} in the following equivalent form:
\begin{eqnarray}
\label{rec-rel-1}
({\textstyle{\partial\over\partial t_{p+1}}}G{\Psi})
= {\lambda}({\textstyle{\partial\over\partial t_{p}}}G{\Psi})
+({\textstyle{\partial\over\partial t_{p}}}G{\theta}^{(-1)})G{\Psi}.
\label{psi-eq-2a}
\end{eqnarray}
Then, substituting $G$ (\ref{transf}), $\Psi$ (\ref{spectrferm}), and
${\theta}^{(-1)}$, derived from the dressing condition \p{locallaxrepr1},
into eq. \p{psi-eq-2a}, the latter becomes
\begin{eqnarray}
\label{rec-rel-2}
&&({\textstyle{\partial\over\partial t_{p+1}}}{\psi}_1)
= \lambda ({\textstyle{\partial\over\partial t_{p}}}{\psi}_1)
+[((\partial^{-1}{\textstyle{\partial\over\partial t_p}}F D{\overline D}
~{\overline F})
-({\textstyle{\partial\over\partial t_p}}F)D{\overline D}\partial^{-1}{\overline F}
\nonumber\\
&&-(\partial^{-1}{\textstyle{\partial\over\partial t_p}}F^* D{\overline F}
g^{-1})(\partial^{-1}g F^* {\overline D}~{\overline F})+
(\partial^{-1}{\textstyle{\partial\over\partial t_p}}F^* D{\overline F}
g^{-1}\partial^{-1}g F^* {\overline D}~{\overline F}))\psi_1]
\end{eqnarray}
where the fermionic derivatives $D$ and ${\overline D}$, entering
the square brackets, act inside these brackets. Eq. \p{rec-rel-2} is satisfied
if and only if eq. \p{spectr} is satisfied, and the latter is obvious if one takes into
account eq. \p{spectrstart} and relation \p{lax^n-iden5} of the Proposition 4.
$~~~~~\blacksquare$

Let us discuss shortly the locality of the isospectral flows in the super--algebraic
Lax--pair representation \p{locallaxrepr}.
The connection ${\cal A}$ \p{laxmatrixz} entering into the Lax operator ${\cal L}$ \p{locallaxrepr1}
is a local functional of the supermatrix--valued superfields $F,~\overline F$ and their derivatives.
It is also known \cite{agnpz} that the matrix $({\Theta}{\lambda}^p E {\Theta}^{-1})_{+}$ is a local
functional. Using \p{transf}, (\ref{d-dbar}--\ref{d})
one can calculate the second term of the Lax representation \p{locallaxrepr}
\begin{eqnarray}
(G^{-1}{\textstyle{\partial\over\partial t_{p}}}G)
=\left(\begin{array}{ccccc}
0 &   0 &  0 &  0 &  0 \\
0 &   0 &  0 &  0 &  0 \\
0 &   0 &  0 &  0 &  0 \\
(L^p)_{D\partial^{-1}} &   0 &  0 &
- res(L^p) &  0 \\
(D~(L^p)_{D\partial^{-1}})&
-(L^p)_{D\partial^{-1}} &  0 &
-(D~ res(L^p))&
- res(L^p)
\end{array}\right)
\label{transf1}
\end{eqnarray}
which is a local functional as well. Thus, all the objects involved into
the Lax--pair representation \p{locallaxrepr} are local, therefore the same is true with respect
to the corresponding isospectral flows.

{}~

{}~

\noindent{\bf Acknowledgments.} A.S. would like to thank the
Laboratoire de Physique de l'ENS Lyon and Institut f\"ur
Theoretische Physik, Universit\"at Hannover, for the hospitality during
the course of this work. This work was partially supported by
the RFBR Grant No. 06-01-00627-a, RFBR-DFG Grant No. 06-02-04012-a,
DFG Grant 436 RUS 113/669-3, the Program for Supporting Leading
Scientific Schools (Grant No. NSh-5332.2006.2), and by the
Heisenberg-Landau Program.


\begin{thebibliography}{**}
\bibitem{ks1}
A.S. Sorin and P.H.M. Kersten, {\it The N=2 supersymmetric
unconstrained matrix GNLS hierarchies}, Lett. Math. Phys. {\bf 60}
(2002) 135,  nlin.SI/0201026.
\bibitem{Delduc-Sorin} F. Delduc and A.S. Sorin,
{\it Recursion operators of the N=2 supersymmetric unconstrained matrix
GNLS hierarchies}, JHEP Proceedings, PrHEP unesp2002, Workshop on Integrable
Theories, Solitons and Duality, 1-6 July 2002, Sao Paulo, Brazil,
nlin.SI/0206037.
\bibitem{bks1}
L. Bonora, S. Krivonos, and A. Sorin, {\it The $N=2$ supersymmetric
matrix GNLS hierarchies}, Lett. Math. Phys. {\bf 45} (1998) 63,
solv-int/9711009.
\bibitem{bks2}
L. Bonora, S. Krivonos, and A. Sorin, {\it Coset approach to the
$N=2$ supersymmetric matrix GNLS hierarchies}, Phys. Lett. {\bf
A240} (1998) 201, solv-int/9711012.
\bibitem{bks}
L. Bonora, S. Krivonos, and A. Sorin, {\it Towards the construction
of $N=2$ supersymmetric integrable hierarchies}, Nucl. Phys. {\bf
B477} (1996) 835, hep-th/9604165.
\bibitem{bs}
L. Bonora and A. Sorin, {\it The Hamiltonian structure of the N=2
supersymmetric GNLS hierarchy}, Phys. Lett. {\bf B407} (1997) 131,
hep-th/9704130.
\bibitem{pop}
Z. Popowicz, {\it The extended supersymmetrization of the
multicomponent Kadomtsev-Petviashvili hierarchy}, J. Phys.
{\bf A29} (1996) 1281, hep-th/9510185.
\bibitem{fk}
A.P. Fordy and P.P. Kulish, {\it Nonlinear Schr\"{o}dinger
equations and simple Lie algebras}, Commun. Math. Phys. {\bf 89}
(1983) 427.
\bibitem{ks2}
P.H.M. Kersten and A.S. Sorin, {\it Bi-Hamiltonian structure of
the $N=2$ supersymmetric ${\alpha}=1$ KdV hierarchy}, Phys. Lett.
{\bf A300} (2002) 397, nlin.SI/0201061.
\bibitem{agnpz}
H. Aratyn, J.F. Gomes, E. Nissimov, S. Pacheva, and A.H. Zimerman,
{\it Symmetry Flows, Conservation Laws and Dressing Approach to
the Integrable Models}, in {\it Integrable Hierarchies and Modern
Physical Theories}, Eds. H. Aratyn and A.S. Sorin, Kluwer Acad.
Publ., Dordrecht/Boston/London, 2001, pg. 243, nlin.SI/0012042.
\bibitem{agz-recent}
H. Aratyn, J.F. Gomes, G.M. de Castro, M.B. Silka, and A.H. Zimerman,
{\it Supersymmetry for integrable hierarchies on loop superalgebras},
J. Phys. {\bf A38} (2005) 9341, hep-th/0508008.

\end{thebibliography}
\end{document}